\documentclass{JHEP3} 

\usepackage{graphicx}
\usepackage{amsmath, amsfonts}
\usepackage{epsfig}

\newcommand{\be}{\begin{equation}}
\newcommand{\ee}{\end{equation}}
\newcommand{\bea}{\begin{eqnarray}}
\newcommand{\eea}{\end{eqnarray}}
\newcommand{\ba}[1]{\begin{array}{#1}}
\newcommand{\ea}{\end{array}}

\title{Electromagnetic signatures
of a strongly coupled anisotropic plasma}

\author{Anton Rebhan, Dominik Steineder\\
Institut f\"{u}r Theoretische Physik, 
Technische Universit\"{a}t Wien,\\
Wiedner Hauptstr. 8--10,
1040 Vienna, Austria\\ 
\\
 \email{rebhana@hep.itp.tuwien.ac.at}\\
  \email{steineder@hep.itp.tuwien.ac.at}}

\abstract{In heavy-ion collisions, quark-gluon plasma is likely to be
produced with sizable initial pressure anisotropy, which may leave
an imprint on electromagnetic observables. In order to model a
strongly coupled anisotropic plasma, we use the AdS/CFT correspondence
to calculate the current-current correlator of a weakly gauged U(1)
subgroup of R symmetry in an $\mathcal N=4$ super-Yang-Mills plasma with a
(temporarily) fixed anisotropy. The dual geometry, obtained
previously by Janik and Witaszczyk, contains a naked singularity
which however permits purely infalling boundary conditions and
therefore the usual definition of a retarded correlator. We obtain
numerical results for the cases of wave vector parallel and
orthogonal to the direction of anisotropy, and we compare with
previous isotropic results. In the (unphysical)
limit of vanishing frequency (infinite time) we obtain a vanishing
DC conductivity for any amount of anisotropy, but the anisotropic
AC conductivities smoothly approach the isotropic case in the
limit of high frequencies.  
We also discuss hard photon production from
an anisotropic plasma and compare with existing hard-loop resummed
calculations.}


\begin{document}

\section{Introduction}
\label{intro}

Information on the earliest stages of heavy-ion collisions and the
production of quark-gluon plasma may be deduced from spectra of
photons and dileptons, because real or virtual photons interact
only weakly with the surrounding strongly coupled matter \cite{Arleo:2004gn,Stankus:2005eq
}. 
Thermal
photon and dilepton production from an isotropic plasma has been studied both in (resummed) perturbation
theory \cite{Aurenche:1998nw,Arnold:2001ms,Arnold:2002ja,Blaizot:2005mj} and in strongly coupled supersymmetric Yang-Mills
plasma by means of gauge/gravity duality \cite{CaronHuot:2006te} (see also \cite{Parnachev:2006ev,Mateos:2007yp,Atmaja:2008mt}).

The angular dependence of photon and dilepton production 
should carry information
on anisotropies present in the plasma prior to isotropization
\cite{Schenke:2006yp,Ipp:2007ng,Mauricio:2007vz,Martinez:2008di,Martinez:2008mc,Bhattacharya:2008mv,Bhattacharya:2009sb},
the details of which still remain to be understood. 
The success of fits of ideal-hydrodynamic simulations \cite{Huovinen:2001cy} suggested
time scales of isotropization of $\,\lesssim 0.7$ fm/c, but later
analysis using viscous hydrodynamics found that thermalization may
be as late as $\,\sim 2$ fm/c, depending significantly on assumptions about the initial conditions \cite{Luzum:2008cw}. 
The recent studies of Ref.\ \cite{Martinez:2010sc,Martinez:2010sd} suggest
that large momentum-space anisotropies may be present for the most part
of the time evolution both at weak and strong coupling.

At parametrically weak coupling, a possible framework for carrying out
systematic calculations in the nonequilibrium case of an initial
momentum space anisotropy is given by the hard-loop effective theory \cite{Mrowczynski:2004kv}
where the backreaction of collective phenomena on the distribution
of hard particles is ignored. The latter is either taken as
stationary and anisotropic \cite{Rebhan:2004ur,Arnold:2005vb,Rebhan:2005re,Arnold:2005ef,Bodeker:2007fw,Arnold:2007cg,Ipp:2010uy}, or free streaming with evolving degree
of anisotropy\cite{Romatschke:2006wg,Rebhan:2008uj,Rebhan:2009ku}. Photon production from Compton-like
and annihilation processes in a stationary anisotropic
quark-gluon plasma has been studied in \cite{Schenke:2006yp}, confirming expectations
of a strong angular dependence. Combined with the time dependence
of the momentum space anisotropy, this may even lead to
yoctosecond pulses of high-energy photons \cite{Ipp:2009ja}. Weak-coupling results also exist for energetic dilepton emission from an anisotropic plasma \cite{Martinez:2008di}, however without inclusion of next-to-leading order effects that become important when dilepton energies are not much higher than temperatures.

In this paper we shall use the AdS/CFT correspondence to study
electromagnetic properties of a strongly coupled anisotropic
plasma in the approximation of a fixed anisotropy. Like in the
weak-coupling situation this should be a reasonable approximation
for processes on time-scales smaller than the isotropization time,
such as the production of hard photons or dilepton pairs.
Following Ref.~\cite{Janik:2008tc} we study
the dual geometry of
a stationary anisotropic $N=4$ supersymmetric Yang-Mills plasma,
which contains a naked anisotropic singularity that however
still permits purely infalling boundary conditions.
We use this setup to generalize the calculation of
U(1) current-current spectral functions of Ref.~\cite{CaronHuot:2006te}
to the anisotropic case. Unlike Ref.~\cite{Janik:2008tc} we refrain
from expansions in the anisotropy parameter, since even the
smallest anisotropy changes the character of the fluctuation
equations. Indeed, the spectral functions are strongly
modified at small frequencies such that the DC conductivity
vanishes. The absence of a horizon at any nonzero anisotropy
leads to the absence of hydrodynamic behavior, which however just
reflects the fact that a stationary anisotropy can only
be a good approximation of the nonequilibrium dynamics
at sufficiently short time scales. 
In fact,
at large frequencies our results for the
spectral function
smoothly approach the isotropic limit. This suggests
that our setup should permit us to
study the effects of temporary anisotropies in a strongly coupled plasma
on electromagnetic observables such as hard photon and dilepton spectra.

\section{The dual geometry}\label{sec:geometry}

There has been much progress recently in studying, numerically,
the gravity dual of the nonequilibrium dynamics corresponding to analogs of
heavy-ion collisions and quark-gluon plasma formation
\cite{Janik:2005zt,Kovchegov:2007pq,Grumiller:2008va,Chesler:2008hg,Lin:2009pn,Gubser:2009sx,Chesler:2009cy,Chesler:2010bi}, building upon and generalizing
the AdS/CFT correspondence
of maximally supersymmetric Yang-Mills
theory at infinite 't Hooft coupling \cite{Aharony:1999ti}. Even though
all these studies indicate very quick transition
to hydrodynamic behavior, estimated to take place at $\,\sim 0.3$ fm/c,
the results of Ref.~\cite{Chesler:2010bi} show significant anisotropies
in the energy-momentum tensor in
the subsequent evolution. This is qualitatively in agreement
with the analysis of Ref.~\cite{Martinez:2010sc,Martinez:2010sd}
studying kinetic theory equations in the presence of momentum space
anisotropies at various amounts of specific shear viscosity with the
result that both at strong and weak coupling one can expect
sizeable pressure anisotropies. Naturally, at weaker coupling
(larger viscosity), the anisotropy of the energy-momentum tensor reaches larger degrees and persists for longer times.

Since the AdS/CFT 
study of electromagnetic spectral functions may be prohibitively
difficult in a time-dependent gravity background involving the formation of (time-dependent) horizons, we shall content ourselves with a stationary anisotropic background obtained previously by Janik and Witaszczyk \cite{Janik:2008tc}. 

\subsection{Review of the Janik-Witaszczyk solution}

According to the AdS/CFT dictionary, the metric of
asymptotically anti-de Sitter space
in Fefferman-Graham coordinates 
\begin{equation}
ds^2=\frac{\gamma_{\mu\nu}(x^\sigma,z)dx^\mu dx^\nu+dz^2}{z^2},
\end{equation} 
where $z$ is the holographic coordinate,
encodes the energy-momentum tensor $\langle T_{\mu\nu}(x^\sigma)\rangle$ of the dual super-Yang-Mills theory
through its asymptotic behavior near the boundary at $z=0$,
\begin{equation}
\gamma_{\mu\nu}(x^\sigma,z)=\eta_{\mu\nu}+z^4 \gamma_{\mu\nu}^{(4)}(x^\sigma)+\mathcal{O}(z^6)
\end{equation}
with
\begin{equation}
\langle T_{\mu\nu}(x^\sigma)\rangle=\frac{N_c^2}{2\pi^2}\gamma_{\mu\nu}^{(4)}(x^\sigma).
\end{equation}
This fixes the boundary condition for the 5-dimensional Einstein equations with a negative cosmological constant
\be
R_{MN}(x,z)=-4g_{MN}(x,z)
\ee
and one can solve for the 5-dimensional metric. Requiring nonsingularity of the resulting geometry would allow one to select physical solutions for the profile of the energy-momentum tensor and its time evolution \cite{Janik:2005zt,Heller:2008fg,Beuf:2009cx}.

However we are interested in a stationary anisotropic energy momentum tensor 
\be
\langle T_{\mu\nu}
\rangle=\textnormal{diag}(\epsilon,P_L,P_T,P_T),
\qquad \langle T^\mu{}_\mu \rangle=0,
\ee 
which we consider as an approximation to the full dynamics at sufficiently short time scales. 
With these boundary conditions the solution for the metric takes the form
\begin{equation}
ds^2=\frac{1}{z^2}\Bigl(-a(z)dt^2+b(z)dx_L^2+c(z)d\mathbf{x}_T^2+dz^2\Bigr).
\label{eq:metric z}
\end{equation}
with
\begin{eqnarray}
a(z)&=&(1+A^2 z^4)^{1/2-\sqrt{36-2B^2}/4}(1-A^2 z^4)^{1/2+\sqrt{36-2B^2}/4}\nonumber\\
b(z)&=&(1+A^2 z^4)^{1/2-B/3+\sqrt{36-2B^2}/12}(1-A^2 z^4)^{1/2+B/3-\sqrt{36-2B^2}/12}\\
c(z)&=&(1+A^2 z^4)^{1/2+B/6+\sqrt{36-2B^2}/12}(1-A^2 z^4)^{1/2-B/6-\sqrt{36-2B^2}/12}.\nonumber
\end{eqnarray}

The parameters $A$ and $B$ are related to the energy density and the pressures 
according to
\begin{eqnarray}\label{epsPLPT}
\epsilon&=&\frac{N_c^2}{2\pi^2}\left[\frac{A^2}{2}\sqrt{36-2B^2}\right],\nonumber\\
P_L&=&\frac{N_c^2}{2\pi^2}\left[\frac{A^2}{6}\sqrt{36-2B^2}-\frac{2A^2 B}{3}\right],\\
P_T&=&\frac{N_c^2}{2\pi^2}\left[\frac{A^2}{6}\sqrt{36-2B^2}+\frac{A^2 B}{3}\right].\nonumber
\end{eqnarray}

The anisotropy of the system is parametrized by the dimensionless parameter $B$. For $B=0$ we obtain the usual anti-de Sitter/Schwarzschild solution in Fefferman-Graham coordinates.
For positive (negative) values of $B$ we obtain an oblate (prolate) anisotropy of the system. $P_L$ and $P_T$ are positive quantities in the range
$-\sqrt{6}\leq B\leq \sqrt{2}$, while in the larger interval $-\sqrt{18}<
B<\sqrt{18}$ one of the pressure components can be made arbitrarily large and negative. The dimensionful parameter $A$, which in the isotropic limit equals $\pi^2 T^2/2$, will be set to 1 in the following discussion.

\begin{table}
\caption{Relation between different anisotropy parameters}
\label{tab:aniso}
\begin{center}
  \begin{tabular}{rccc}
\hline
  $B$ & $\xi$ & $P_L/\epsilon$ & $P_T/\epsilon$ \\
\hline
$-4$ & - & 3 & $-1$ \\
$-\sqrt6$ & $-1$ & 1 & 0\\
   $-1$ & $-0.69675$ & 0.5620 & 0.2190 \\
   $-0.1$& $-0.1160$ 
                & 0.35556 & 0.3222 \\
   0  &  0   & 1/3 & 1/3  \\
   0.1 & 0.1355 
                & 0.3111 & 0.3444\\
   1 & 4.8102 & 0.1047 & 0.4477 \\
$\sqrt2$ & $\infty$ & 0 & 1/2 \\
$\sqrt{12}$ & - & $-1$ & 1 \\
\hline 
 \end{tabular}
\end{center}
\end{table}

In Table \ref{tab:aniso} we display the dependence of the pressure
components normalized to the energy density as a function of $B$
for a number of values used further below in numerical evaluations.
Also given is the connection with the anisotropy parameter $\xi$
introduced in Ref.\ \cite{Romatschke:2003ms} through one-dimensional
deformations of isotropic particle distributions $f_{iso}$ in momentum
space,
\begin{equation}\label{fiso}
 f(\mathbf{p})=\mathcal{N}(\xi)f_{iso}\Big(\sqrt{p^2+\xi p_L^2}\Big),
\end{equation}
with $\mathcal{N}(\xi)$ some normalization factor (e.g., $\mathcal N=\sqrt{1+\xi}$ if the number density is to be kept fixed),
and the corresponding energy-momentum tensor in kinetic theory
\begin{equation}
 T^{\mu\nu}=N_{\rm eff.}\int\frac{d^3\mathbf{p}}{(2\pi)^3}p^\mu p^\nu f(\mathbf{p}).
\end{equation}
(This of course covers only the range where both $P_L$ and $P_T$ remain
positive.)

Using the analytic expressions for the corresponding $\epsilon$, $P_L$ and $P_T$ of
Ref.\ \cite{Rebhan:2008uj},
the relation between $\xi$ and $B$ is given by
\begin{equation}\label{Bvsxi}
 \frac{\sqrt{36-2B^2}+2B}{\sqrt{36-2B^2}-4B}=\frac{\xi-1}{2}+\frac{\xi}{(\xi+1)\xi^{-1/2}{\,\textnormal{atan}\,\xi^{1/2}}-1}.
\end{equation}
For small anisotropies $\xi= \frac54 B+O(B^2)$.

\subsection{The appearance of naked singularities}

\begin{figure}
\centerline{\includegraphics[width=0.55\textwidth]{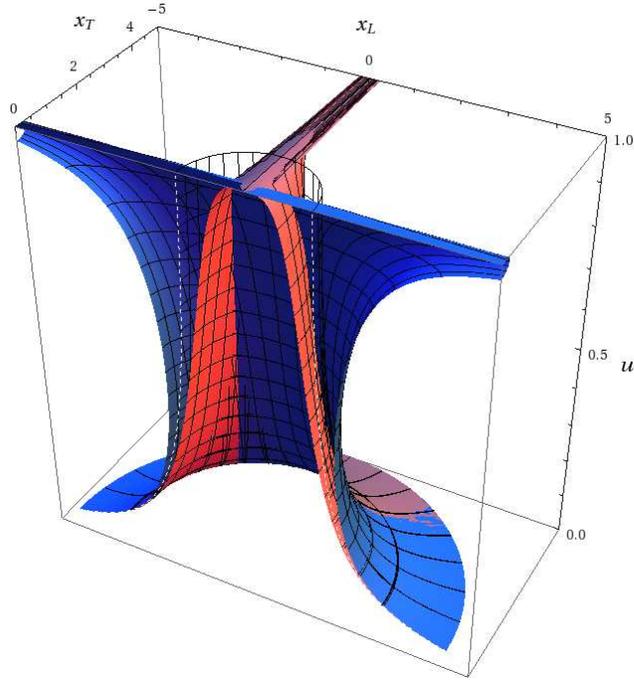}}
\caption{Asymptotically spherical congruences of (holographically) radial light-like geodesics which get deformed into ellipsoids as they approach the singularity at $u=1$ in units where $A=1$. The blue (darker) surface corresponds to prolate anisotropy $B=-\sqrt6$, the red (lighter) surface to oblate anisotropy $B=\sqrt2$, and the transparent mesh to the isotropic case $B=0$. 
Here $x_T$ and $x_L$ correspond to the transverse and longitudinal spatial extents of the ellipsoids, which degenerate into an infinite disk or line for oblate or prolate anisotropy, respectively. (Note that $x_T$ is a radial variable.)
\label{fig:exhib}}
\end{figure}

For later convenience we write the metric in terms of $u=z^2$: 
\begin{equation}
ds^2=g_{tt}(u)dt^2+g_{LL}(u)dx_L^2+g_{TT}(u)d\mathbf{x}_T^2+g_{uu}(u)du^2, 
\label{eq:metric u}
\end{equation}
\begin{eqnarray}
g_{tt}(u)&=&-\frac{1}{u}(1+A^2 u^2)^{1/2-\sqrt{36-2B^2}/4}(1-A^2 u^2)^{1/2+\sqrt{36-2B^2}/4}\nonumber\\
g_{LL}(u)&=&\frac{1}{u}(1+A^2 u^2)^{1/2-B/3+\sqrt{36-2B^2}/12}(1-A^2 u^2)^{1/2+B/3-\sqrt{36-2B^2}/12}\nonumber\\
g_{TT}(u)&=&\frac{1}{u}(1+A^2 u^2)^{1/2+B/6+\sqrt{36-2B^2}/12}(1-A^2 u^2)^{1/2-B/6-\sqrt{36-2B^2}/12}\\
g_{uu}(u)&=&\frac{1}{4u^2}.\nonumber
\end{eqnarray}

The geometry corresponding to this metric is pathological in the sense that a naked singularity appears whenever $B$ does not vanish. For instance, the induced metric at constant $t$ and $u=1/A$ is degenerate,
\be
g_{LL}g_{TT}^2\propto (1-A^2 u^2)^{\left[6-\sqrt{36-2B^2}\right]/4}.
\ee
Three-dimensional space degenerates at $u=1/A$ 
into a two-dimensional sheet when $B>0$, and into a one-dimensional line when
$B<0$. This is illustrated in Fig.~\ref{fig:exhib} in terms of
asymptotically $(u\to0)$ spherical congruences of holographically radial light-like geodesics. At finite $u$ these are deformed into ellipsoids, which degenerate at $u=1/A$.

However, in \cite{Janik:2008tc} it was noted that it is nevertheless possible to define purely ingoing and outgoing boundary conditions at the naked singularity,\footnote{But the expansion in small $B$ performed in \cite{Janik:2008tc} is not really allowed, because the character of the singularities in the equations of motion changes with the anisotropy parameter. We thank Karl Landsteiner for pointing this out to us.} which 
gives the possibility to define retarded correlation functions along
the lines of Ref.\ \cite{Son:2002sd}. 
In the following we shall do so for current-current correlators of a U(1) current selected from the $R$-charge currents and assumed to be weakly gauged and thus coupled to electromagnetism.

\section{Current-current correlators and photon 
production}
\label{sec:photon and dilepton}
Photons and leptons produced in an early stage after the heavy ion collision interact only very weakly with the medium and can leave the plasma without any rescattering. Therefore the production of photons and dileptons in a strongly coupled plasma is an interesting observable to study the far from equilibrium properties. 
In isotropic $\mathcal{N}=4$ SYM plasmas the emission spectra of photons and leptons at strong and weak coupling have been studied in \cite{CaronHuot:2006te}.
In the following we shall closely follow this paper and generalize to the above anisotropic background.

\subsection{Basic facts}

The rate of photons produced per unit time and per unit volume is\footnote{We use capital letter to denote 4 vectors and lower case letters for the absolute value of 3 vectors. The Minkowski metric is defined as $\eta_{\mu\nu}=\textnormal{diag}(-1,1,1,1)$.} 
\begin{equation}
d\Gamma_\gamma=\frac{d^3 k}{(2\pi)^3}\frac{i e^2}{2|\mathbf{k}|}\eta^{\mu\nu}C_{\mu\nu}^{<}(K)|_{k^0=|\mathbf{k}|},
\label{eq:photon production}
\end{equation}
where 
\begin{equation}
C_{\mu\nu}^{<}(K)=\int d^4 X e^{-i K\cdot X}\langle J_\mu^{EM}(0)J_\nu^{EM}(X)\rangle
\end{equation}
is the Wightman function of electromagnetic currents.
Any correlator of the conserved current can be expressed in terms of the spectral function, which is defined as
\begin{equation}
\chi_{\mu\nu}(K)=\int d^4 X e^{-i K\cdot X}\langle [J_\mu^{EM}(0),J_\nu^{EM}(X)]\rangle.
\end{equation}
Later we will determine this spectral function by making use of 
\begin{equation}
\chi_{\mu\nu}(K)=-2\textnormal{Im} C_{\mu\nu}^{ret}(K).
\end{equation}
The Wightman function is related to the spectral function by 
\begin{equation}
 C^<_{\mu\nu}(K)=-i\chi_{\mu\nu}(K)\phi(K),
\end{equation}
where $\phi$ reduces to the Bose-Einstein distribution in thermal equilibrium. 

The dilepton production takes place via an intermediate virtual photon and the rate is given by
\begin{equation}
d\Gamma_{l\bar{l}}=\frac{d^4 K}{(2\pi)^4}\frac{i e^2 e_l^2(-K^2-4m^2)^{1/2}(-K^2+2m^2)}{6\pi|K^2|^{5/2}}\theta(k^0)\theta(-K^2-4m^2)\eta^{\mu\nu}C_{\mu\nu}^{<}(K),
\label{eq:dilepton production}
\end{equation}
where $e_l$ is the charge of the lepton. Here the Wightman function must be evaluated for time-like momenta.
 
\subsection{Introduction of photons and leptons}
After the brief recapitulation of photon and dilepton production rates in thermal media, we quickly review the possibilities to couple electromagnetism to the $\mathcal{N}=4$ SYM theory, which consists of $SU(N_c)$ gauge bosons, four Weyl fermions $\psi_p$ and six real scalars $\phi_{pq}=-\phi_{qp}$, all transforming in the adjoint representation of $SU(N_c)$. There is also an anomaly free global $SU(4)$ $R$-symmetry present, under which the fermions transform in the $\mathbf{4}$ and the scalars in the $\mathbf{6}$. We can take a $U(1)$ subgroup of the $SU(4)$ $R$-symmetry associated with a $U(1)$ gauge field coupled to the conserved current. By doing so we are able to model the electromagnetic interactions. In principle it is possible to take any linear combination of Cartan subalgebra generators to embed the $U(1)$ in the $SU(4)$ $R$-symmetry group (for details see \cite{CaronHuot:2006te}). We will choose $t^3=\textnormal(1/2,-1/2,0,0)$ to be the generator of the $U(1)$ such that two of the Weyl fermions have charge $\pm 1/2$ and two complex scalars have charge $1/2$. The conserved current is
\begin{equation}
J^{EM}_\mu=\frac{1}{e}\frac{\delta S_{int}}{\delta A^\mu}=\frac{1}{2}\Big(\psi_1^{a\dagger}\bar{\sigma}_\mu\psi_1^a-\psi_2^{a\dagger}\bar{\sigma_\mu}\psi_2^a+\sum_{p=3,4}\phi_{1p}^{a\dagger}(-i\overrightarrow{D}_\mu+i\overleftarrow{D}_\mu)\phi_{1p}^a\Big), 
\end{equation} 
where $a$ is the $SU(N_c)$ group index and the covariant derivative $D_\mu$ involves the $SU(N_c)$ gauge fields as well as the $U(1)$ electromagnetic vector potential $A_\mu$. Because we are only interested in the leading order terms in the electromagnetic coupling $e$, it is sufficient to treat the electromagnetic interaction as being linear in $A_\mu$. Then we can consistently ignore the electromagnetic vector potential in the covariant derivative acting on the scalar. In order to also add weakly coupled leptons $l$ with charge $e_l$ and mass $m$, the Lagrangian is extended to
\begin{equation}
\mathcal{L}=\mathcal{L}_{SYM}+\mathcal{L}_{int}-\frac{1}{4}F_{\mu\nu}^2-\bar{l}(\displaystyle{\not}D+m)l \qquad \textnormal{with }\mathcal{L}_{int}=eJ_\mu^3 A^\mu,
\end{equation} 
where $J_\mu^3$ is the $t^3$ component of the $R$-current.

In the gauge/gravity setup, only the SYM part will be realized dynamically, but we can calculate current-current correlators to leading order in the electromagnetic coupling and to all orders in the SYM coupling.

\subsection{Tensor structure of anisotropic correlators}

For an anisotropic medium, the tensor structure of the current-current correlator is more complicated than in the isotropic finite temperature case.
Because the electromagnetic current is conserved it must satisfy Ward identities, which implies that any $C_{\mu\nu}(K)\sim P_{\mu\nu}f(K)$ with $P_{\mu\nu}=\eta_{\mu\nu}-K_\mu K_\nu/K^2$. 

We begin with the case when the wave vector is pointing in the direction of the anisotropy denoted by $\mathbf{n}=n\mathbf{e}_L$. Then it is sufficient to introduce longitudinal and transverse projectors
\begin{eqnarray}
P^T_{00}=P^T_{i0}=0, \qquad P^T_{ij}=\delta_{ij}-\frac{k_i k_j}{k^2},\qquad P^L_{\mu\nu}=P_{\mu\nu}-P^T_{\mu\nu}.
\end{eqnarray}
For the correlators we then find
\begin{equation}
C_{\mu\nu}(K)=P^T_{\mu\nu}\Pi^T(K)+P^L_{\mu\nu}\Pi^L(K).
\end{equation}
This is exactly the same structure one has in isotropic systems. 

If we choose the wave vector to point in a perpendicular direction with respect to the anisotropy $\mathbf{k}=k\mathbf{e_1}$, 
we need one further tensorial structure,
\begin{equation}
P^2_{00}=P^2_{i0}=0, \qquad P^2_{ij}=\delta_{ij}-\frac{k_i k_j}{k^2}-\frac{n_i n_j}{n^2}
\end{equation}
\begin{equation}
P^L_{00}=P^L_{i0}=0, \qquad P^L_{ij}=\frac{n_i n_j}{n^2}
\end{equation}
\begin{equation}
P^1_{\mu\nu}=P_{\mu\nu}-P^2_{\mu\nu}-P^L_{\mu\nu}.
\end{equation}
As a consequence the correlator is then specified by three scalar functions 
\begin{equation}
C_{\mu\nu}(K)=P^1_{\mu\nu}\Pi^1(K)+P^2_{\mu\nu}\Pi^2(K)+P^L_{\mu\nu}\Pi^L(K).
\end{equation}

With a generic orientation of the wave vector, more structure functions
would come into the play. We shall restrict our attention to the two
extreme cases of wave vector parallel and orthogonal to the anisotropy
direction. It is plausible that the generic case will interpolate smoothly between
those two.

\subsection{Equations of motion and asymptotic solution}
To obtain the retarded correlator we have to solve the equations of motion of a gauge field in the geometry described by (\ref{eq:metric u}), which are given by 
\begin{equation}
\partial_A(\sqrt{-g}g^{AC}g^{BD}F_{CD})=0.
\end{equation}
At first we consider the case when the wave vector points in the direction of the anisotropy ($\mathbf{k}=k_L\mathbf{e_L}$). Then the equation of motion for $E_T=\omega A_T$ is
\begin{equation}
E''_T+\frac{\partial_u(\sqrt{-g}g^{uu}g^{TT})}{\sqrt{-g}g^{uu}g^{TT}}E'_T-\frac{g^{tt}\omega^2+g^{LL}k_L^2}{g^{uu}}E_T=0,
\label{eq:EoM ET1}
\end{equation}
where primes denote derivatives with respect to $u$ and the dependence of $u$ is suppressed everywhere. For the longitudinal electric field $E_L=k_L A_t+\omega A_L$ we find
\begin{eqnarray}\label{ELodepar}
E''_L+\frac{(g^{tt})^2\partial_u(\sqrt{-g}g^{uu}g^{LL})\omega^2+(g^{LL})^2\partial_u(\sqrt{-g}g^{uu}g^{tt})k_L^2}{\sqrt{-g}g^{uu}g^{tt}g^{LL}(g^{tt}\omega^2+g^{LL}k_L^2)}E'_L&&\nonumber\\
-\frac{g^{tt}\omega^2+g^{LL}k_L^2}{g^{uu}}E_L&=&0.
\end{eqnarray}

If the wave vector points in a direction perpendicular to the anisotropy, we obtain three different differential equations. Due to symmetry we are free to choose $\mathbf{k}=k_1 \mathbf{e}_1$ and then find
\begin{eqnarray}
E''_2+\frac{\partial_u(\sqrt{-g}g^{uu}g^{TT})}{\sqrt{-g}g^{uu}g^{TT}}E'_2-\frac{g^{tt}\omega^2+g^{TT}k_1^2}{g^{uu}}E_2&=&0\\
E''_L+\frac{\partial_u(\sqrt{-g}g^{uu}g^{LL})}{\sqrt{-g}g^{uu}g^{LL}}E'_L-\frac{g^{tt}\omega^2+g^{TT}k_1^2}{g^{uu}}E_L&=&0
\label{ELodeperp}
\end{eqnarray}
for the two modes transverse to the wave vector and 
\begin{eqnarray}
E''_1+\frac{(g^{tt})^2\partial_u(\sqrt{-g}g^{uu}g^{TT})\omega^2+(g^{TT})^2\partial_u(\sqrt{-g}g^{uu}g^{tt})k_1^2}{\sqrt{-g}g^{uu}g^{tt}g^{TT}(g^{tt}\omega^2+g^{TT}k_1^2)}E'_1&&\nonumber\\
-\frac{g^{tt}\omega^2+g^{TT}k_1^2}{g^{uu}}E_1&=&0
\end{eqnarray}
for the electric field along the direction of the wave vector. These equations are quite lengthy if the explicit form of the metric coefficients is inserted, therefore we will not do so here. 

For all of these equations, we can use a Frobenius ansatz near the boundary ($u=0$) and find the characteristic exponents to be $0$ and $1$. Close to the naked singularity 
(which appears exactly where the horizon is in the isotropic case)
the differential equations have the following form,
\begin{equation}
\frac{d^2}{du^2}\phi+\frac{C_1}{(1-u)}\frac{d}{du}\phi+\frac{\omega^2 C_2}{(1-u)^\alpha}\phi=0
\end{equation}
with $\alpha=(2+\sqrt{36-2B^2})/4\leq2$. For isotropic systems $\alpha=2$ and a Frobenius ansatz is still possible about $u=1$. In that case we find the characteristic exponents $\pm i\omega/\sqrt{8}$ near the horizon and we can easily define ingoing boundary conditions. For nonvanishing anisotropy $\alpha<2$ and we can perform a coordinate transformation $x=(1-u)^{(2-\alpha)}$ in order to find appropriate boundary conditions at the naked singularity\footnote{Thanks to Karl Landsteiner for clarifying this point.}. Then the equation of motion is given by
\begin{equation}
\frac{d^2}{dx^2}\phi+\frac{\beta}{x}\frac{d}{dx}\phi+\frac{\gamma^2}{x}\phi=0
\end{equation}
where
\begin{equation}
\beta=1-\frac{C_1+1}{2-\alpha}\qquad \gamma^2=\frac{C_2\omega^2}{(2-\alpha)^2}.
\end{equation}
The solution to this differential equation can be written as
\begin{equation}
\phi(u)\sim(1-u)^{(2-\alpha)(1-\beta)/2}\textnormal{H}^{(1,2)}_{1-\beta}(2\gamma(1-u)^{(2-\alpha)/2}),
\end{equation} 
where the Hankel function of the second kind $\mathrm H^{(2)}_\nu$ represents ingoing boundary conditions, which we will use for our numerical studies later on. 
\subsection{Spectral functions at strong coupling}
In order to find the on-shell boundary action, we start from
the five-dimensional Maxwell action, which is given by
\begin{equation}
 S_{5D}=-\frac{1}{4 g_B^2}\int \sqrt{-g}g^{AC}g^{BD}F_{AB}F_{CD}
\end{equation}
with $g_B=16\pi^2 R/N_c^2$ and $R$ the AdS radius.
Choosing the gauge 
$A_u=0$ 
we obtain the on-shell boundary term
\begin{eqnarray}
 S_B&=&-\frac{1}{2g_B^2}\int_{u\rightarrow0}\sqrt{-g}g^{uu}\Big(g^{tt}A'_t(K,u) A_t(-K,u)\nonumber\\
&&+g^{LL} A'_L(K,u) A_L(-K,u)+g^{TT}\mathbf{A}'_T(K,u)\mathbf{A}_T(-K,u)\Big)
\label{eq:S_B with A}
\end{eqnarray}
Considering a wave vector pointing into the direction of the anisotropy first, we obtain the relation
\begin{equation}
 \omega g^{tt} A'_t(K,u)-k g^{LL} A'_L(K,u)=0,
\end{equation}
which follows directly from the equations of motion. Inserting this into (\ref{eq:S_B with A}) and rewriting the result in terms of electric fields $E_L=\omega A_L+k A_t$ and $\mathbf{E}_T=\omega \mathbf{A}_T$, the boundary term of the action becomes
\begin{eqnarray}
 S_B&=&-\frac{1}{2g_B^2}\int_{u\rightarrow0}\sqrt{-g}g^{uu}\Bigg(\frac{g^{tt}}{\omega^2{g^{tt}}/{g^{LL}}+k^2}E'_L(K,u) E_L(-K,u)\nonumber\\
&&+\frac{g^{TT}}{\omega^2}\mathbf{E'}_T(K,u)\mathbf{E}_T(-K,u)\Bigg).
\end{eqnarray}
The transverse correlator is defined as
\begin{equation}
 C_{TT}(K)=\frac{\delta^2 S_B}{\delta A_T(K) \delta A_T(-K)}=\frac{\omega^2 \delta^2 S_B}{\delta E_T(K) \delta E_T(-K)}.
\end{equation}
Applying the Lorentzian AdS/CFT prescription \cite{Son:2002sd} and inserting the explicit form of the metric coefficients we find
\begin{equation}
 C_{TT}(K)=\Pi_{TT}(K)=-\frac{2}{g_B^2}\lim_{u\rightarrow 0}\frac{E'_T(K,u)}{E_T(K,u)}.
\end{equation}
After a similar computation for the longitudinal correlator we find
\begin{equation}
\Pi_L(K)=-\frac{2}{g_B^2}\lim_{u\rightarrow 0}\frac{E'_L(K,u)}{E_L(K,u)}.
\end{equation}

When the wave vector is in the 1-direction we obtain three scalar functions which are given by
\begin{equation}
\Pi_a(K)=-\frac{2}{g_B^2}\lim_{u\rightarrow 0}\frac{E'_a(K,u)}{E_a(K,u)},
\end{equation}
with $a=1,2,L$. Notice that the equations of motion for $E_L$ differ for wave vector perpendicular and parallel to the direction of anisotropy, cf.\ (\ref{ELodepar}) and (\ref{ELodeperp}). 

\section{Numerical results}

\subsection{Wave vector parallel to anisotropy direction}

\begin{figure}
\begin{center}
  \includegraphics[width=0.55\textwidth]{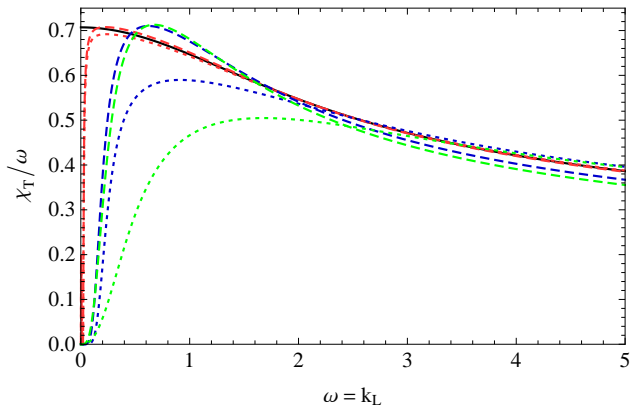}
\end{center}
\caption{Transverse contribution to spectral density for light-like momenta with values for the anisotropy parameter $B=0$ (black), $B=0.1$ (red, dashed), $B=-0.1$ (red, dotted), $B=1$ (blue, dashed), $B=-1$ (blue, dotted), $B=\sqrt{2}$ (green, dashed) and $B=-\sqrt{6}$ (green,dotted).
The dimensionful parameter $A$, which equals $\pi^2T^2/2$ in the isotropic case $B=0$, has been set to unity.}
\label{fig:Light_TL}
\end{figure}

First we discuss the form of the spectral function for a wave vector parallel to the anisotropy direction. When we consider light-like momenta only the transverse contribution to the spectral density is nonvanishing. The results for different anisotropy parameters are shown in Fig.~\ref{fig:Light_TL}. The black line corresponds to the isotropic case and coincides with the spectral density presented in \cite{CaronHuot:2006te} after the correct normalization is chosen. For nonvanishing anisotropy we notice a qualitative difference for small frequencies, namely all spectral functions tend to zero faster than $\omega$, which is in stark contrast to the isotropic situation. However for small values of $B$ the curve quickly rises and then settles close to the isotropic one. For increasing anisotropy the spectral densities deviate more strongly from the isotropic result, in particular at small frequencies.

Since small frequencies correspond to long time scales, the approximation of a stationary anisotropy is bound to break down in this limit. On the other hand, at larger frequencies our analysis should be able to map the nonequilibrium situation in the form of a snapshot, so it is reassuring that there the effects of the anisotropy connect smoothly with the isotropic limit. Given that the appearance of the naked singularity is not a small modification as it changes the character of the differential equations, this is not completely obvious a priori.

\begin{figure}
 \begin{center}
\includegraphics[width=0.48\textwidth]{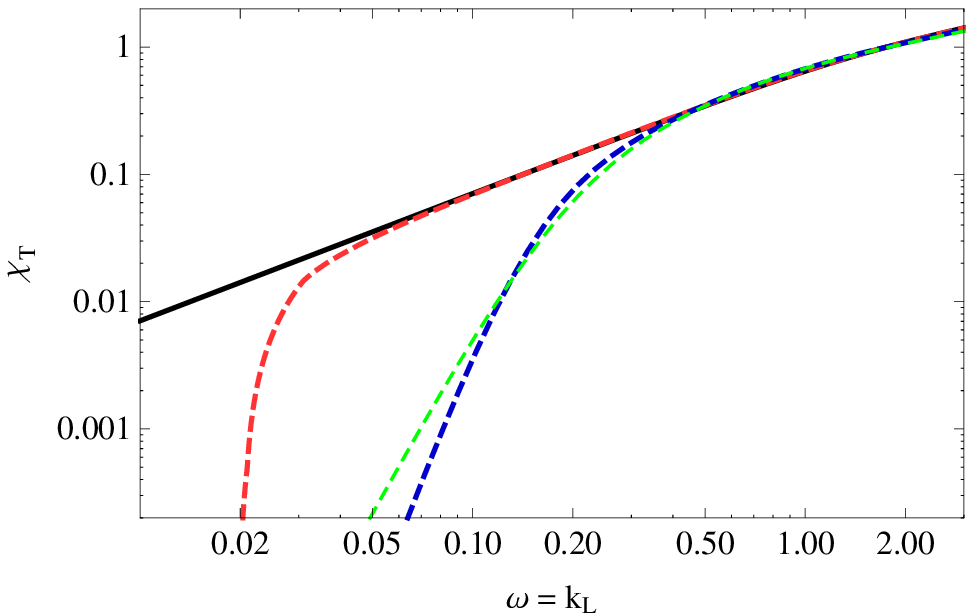}\quad
\includegraphics[width=0.48\textwidth]{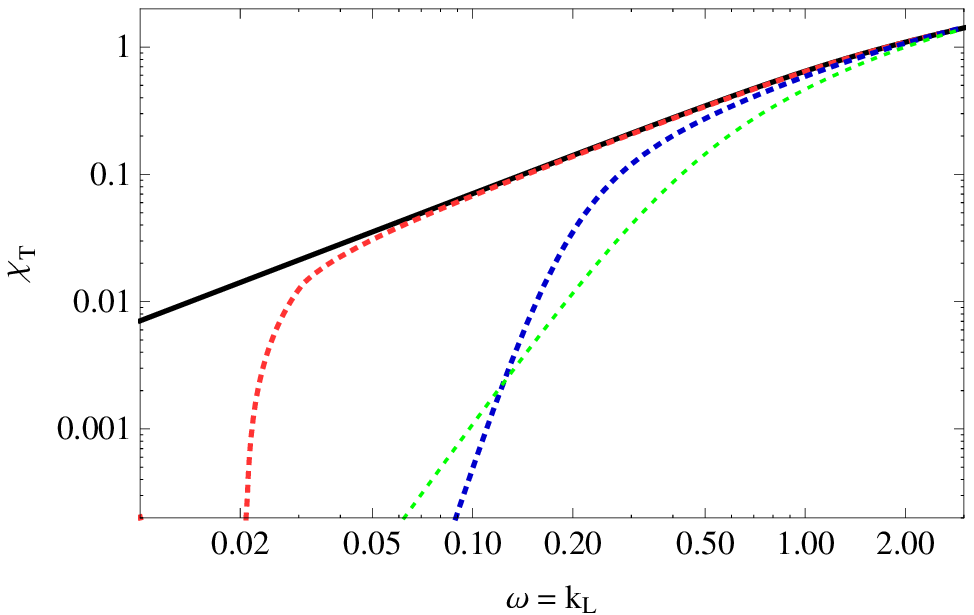}
\caption{Double logarithmic plot to estimate the deviation from the linear behavior of the spectral function with respect to $\omega$. In the left panel we consider positive anisotropies and in the right panel negative ones. Color coding as in Fig. \protect\ref{fig:Light_TL}
}
\label{fig:LogLog}
 \end{center}
\end{figure}

Since the low frequency limit seems to be unphysical,
the question arises down to what value of $\omega$ we might trust this calculation. It makes sense to assume that the lower bound of the frequency will depend on the anisotropy of the system. For larger anisotropies our assumption of a time invariant background may break down earlier. A possible hint to estimate the range of validity of the calculation presented here can come from considering the deviation from $\chi\propto\omega$ as is done in Fig. \ref{fig:LogLog}. We see that even for the most extreme anisotropy parameters this suggests our calculation to be valid down to $\omega\sim 1$ (in units where $A=1$).

\begin{figure}
 \begin{center}
  \includegraphics[width=0.48\textwidth]{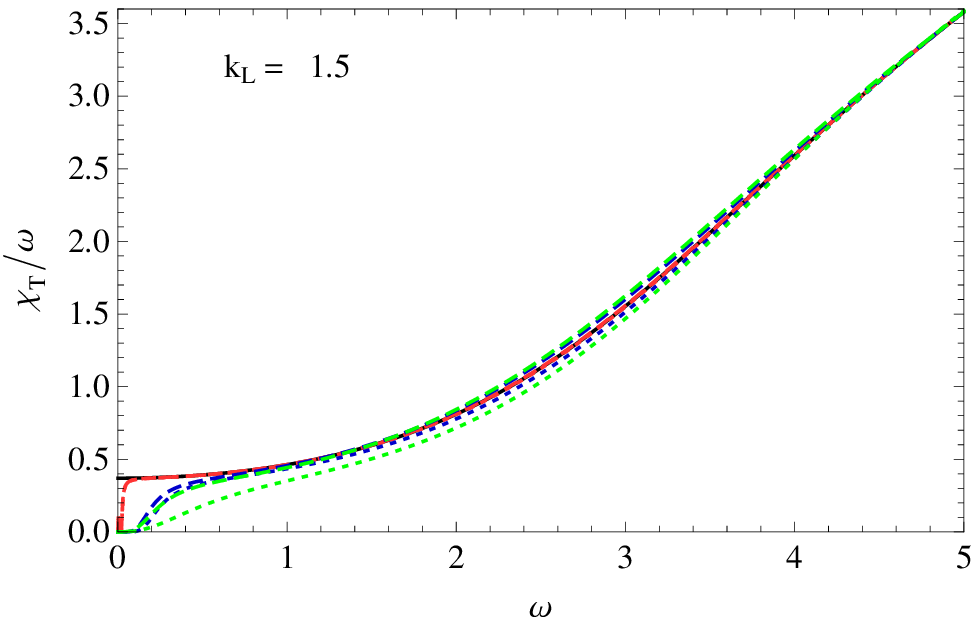}\quad%
\includegraphics[width=0.48\textwidth]{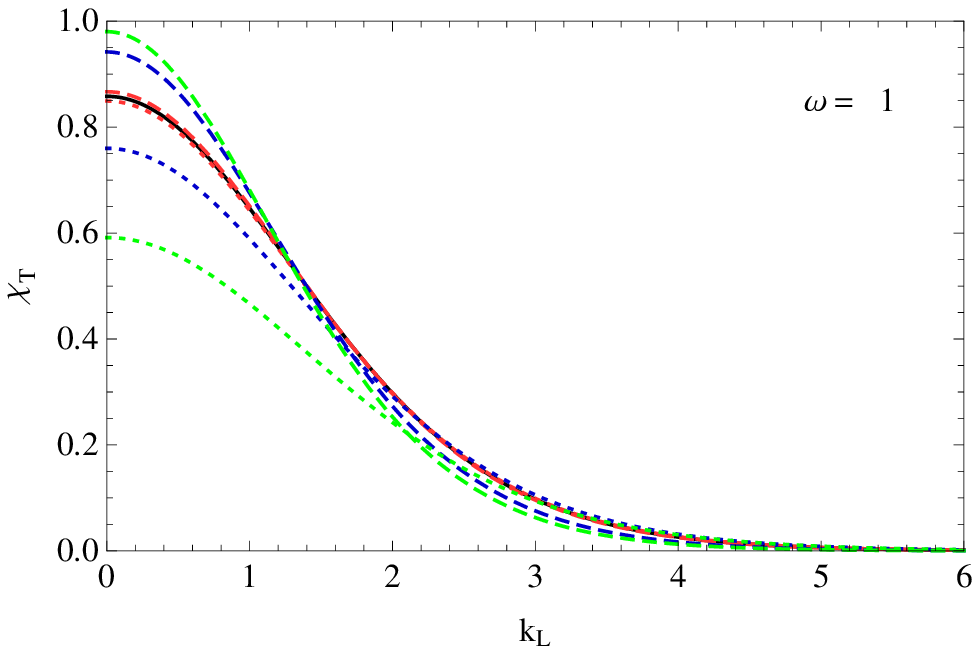}
\caption{Transverse spectral function for $k_L$ fixed (left) and $\omega$ fixed (right). Color coding as in Fig. \protect\ref{fig:Light_TL}}
\label{fig:Momenta_TL}
 \end{center}
\end{figure}

In Fig. \ref{fig:Momenta_TL} the form of the transverse part of the spectral function is shown for $k_L$ fixed and $\omega$ fixed, respectively. In the left panel the difference in the small frequency behavior between isotropic and anisotropic results is again obvious.

\begin{figure}
 \begin{center}
  \includegraphics[width=0.48\textwidth]{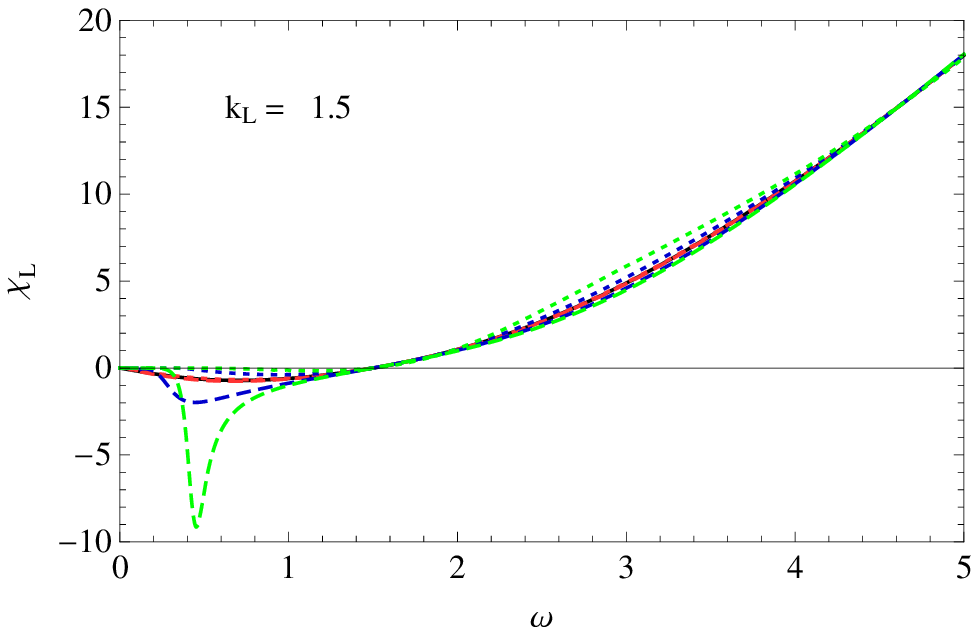}\quad%
\includegraphics[width=0.48\textwidth]{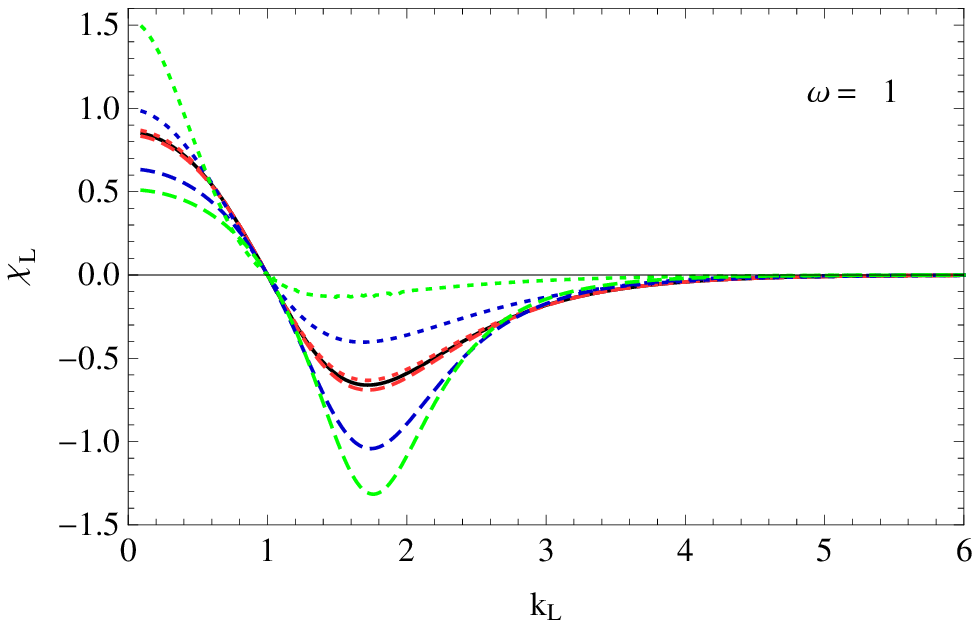}
\caption{Longitudinal spectral function for $k_L$ fixed (left) and $\omega$ fixed (right). Color coding as in Fig. \protect\ref{fig:Light_TL}}
\label{fig:Momenta_LL}
 \end{center}
\end{figure}

As already mentioned the longitudinal part of spectral density vanishes for light-like momenta. This is still the case if we turn on the anisotropy. In Fig. \ref{fig:Momenta_LL} we see that the spectral function is negative for space-like momenta, vanishes for light-like momenta and becomes positive for time-like momenta.\footnote{It turned out to be hard to solve the differential equations in the longitudinal cases numerically. While the default setting of Mathematica's NDSolve produces error messages at some data points, but matches with most of the other methods whenever no error occurs, the so called ``StiffnessSwitching'' method produces no errors at all but sometimes produces very different results from the other methods. We circumvented the difficulties by using the default method and computing many data points, such that we can safely neglect all the data points producing an error message. The final plots do not change when the number of data points is further increased.}

\subsection{Wave vector perpendicular to anisotropy direction}

\begin{figure}
\begin{center}
  \includegraphics[width=0.55\textwidth]{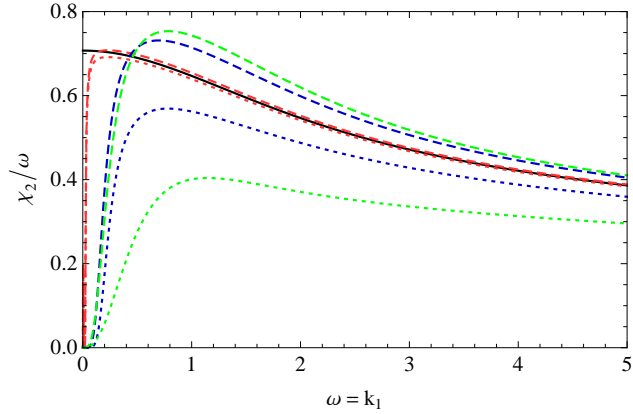}
\caption{Part of spectral density perpendicular to $k_1$ and to the anisotropy direction for light-like momenta. Color coding as in Fig. \protect\ref{fig:Light_TL}}
\label{fig:Light_2T}
\end{center}
\end{figure}

\begin{figure}
 \begin{center}
  \includegraphics[width=0.48\textwidth]{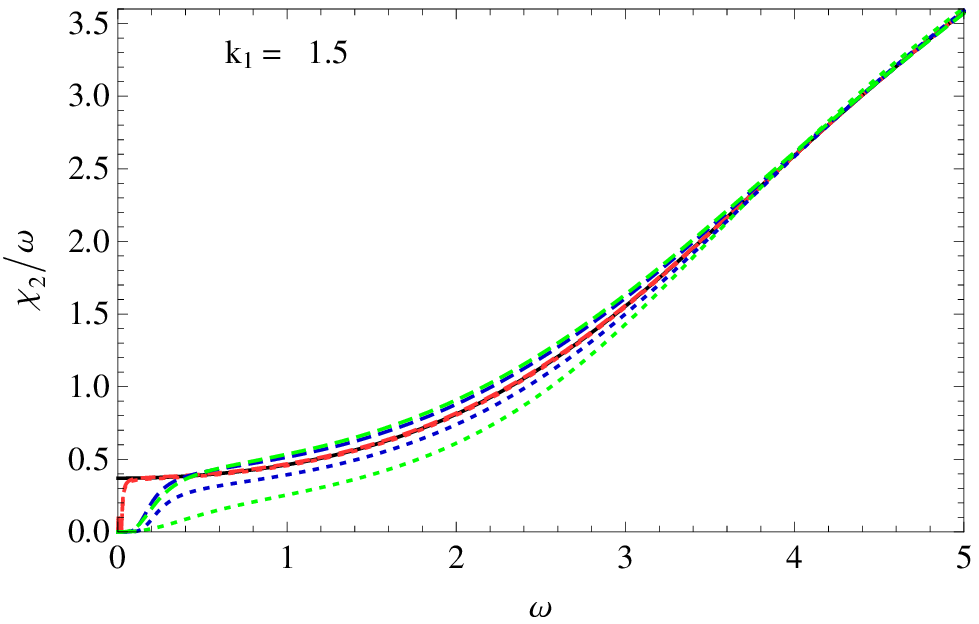}\quad%
\includegraphics[width=0.48\textwidth]{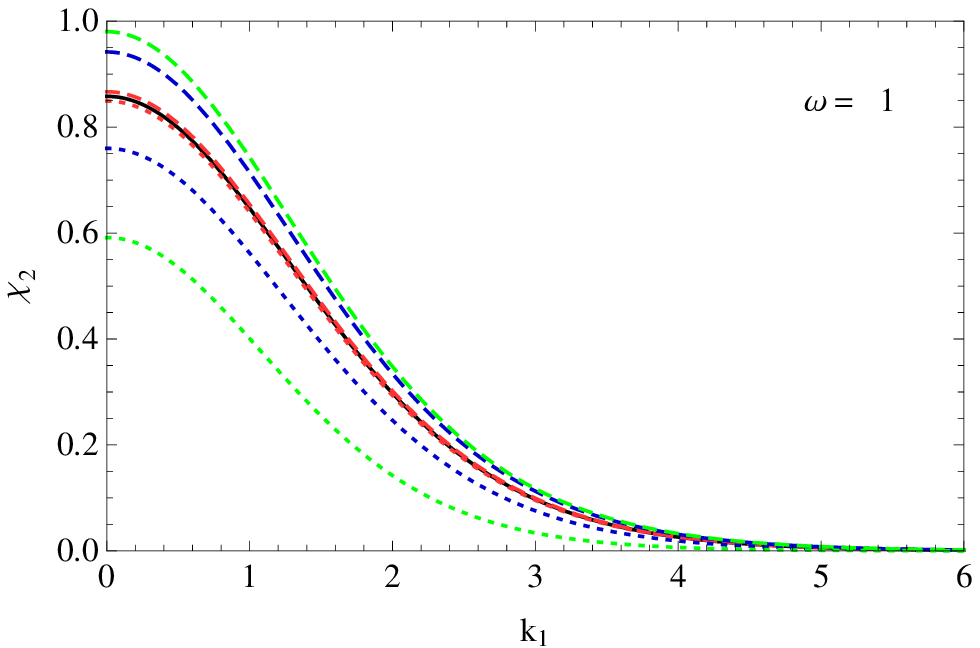}
\caption{Contribution to spectral function which is perpendicular to the wave vector and the anisotropy direction for $k_1$ fixed (left) and $\omega$ fixed (right). Color coding as in Fig. \protect\ref{fig:Light_TL}}
\label{fig:Momenta_2T}
 \end{center}
\end{figure}

Next we consider a wave vector pointing in the $1$-direction. Then there is a mode which is both, perpendicular to the wave vector and the anisotropy direction. This mode can be compared to the transverse mode before and we see that the behavior for small frequencies and wave vectors is quite similar. The spectral function is larger for oblate anisotropy and smaller for prolate. However, while this is still true for larger frequencies and momenta in the present case (see Fig. \ref{fig:Light_2T} and \ref{fig:Momenta_2T}), the opposite was true for $\chi_T$, where the behavior changed at some intermediate frequency or momentum.

\begin{figure}
\begin{center}
  \includegraphics[width=0.55\textwidth]{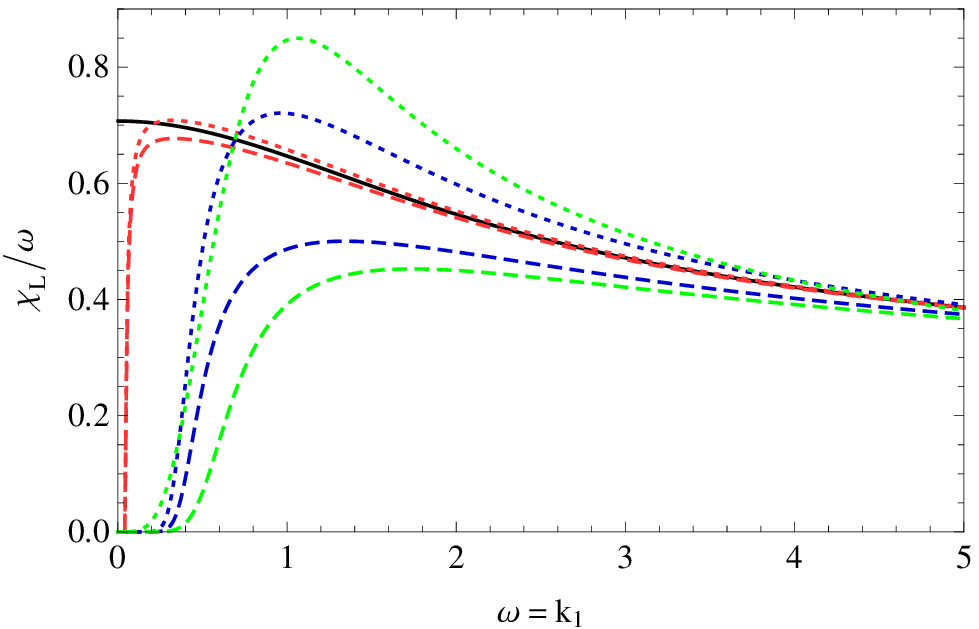}
\caption{Spectral function along the anisotropy direction for light-like momenta. Color coding as in Fig. \protect\ref{fig:Light_TL}}
\label{fig:Light_LT}
\end{center}
\end{figure}

\begin{figure}
 \begin{center}
  \includegraphics[width=0.48\textwidth]{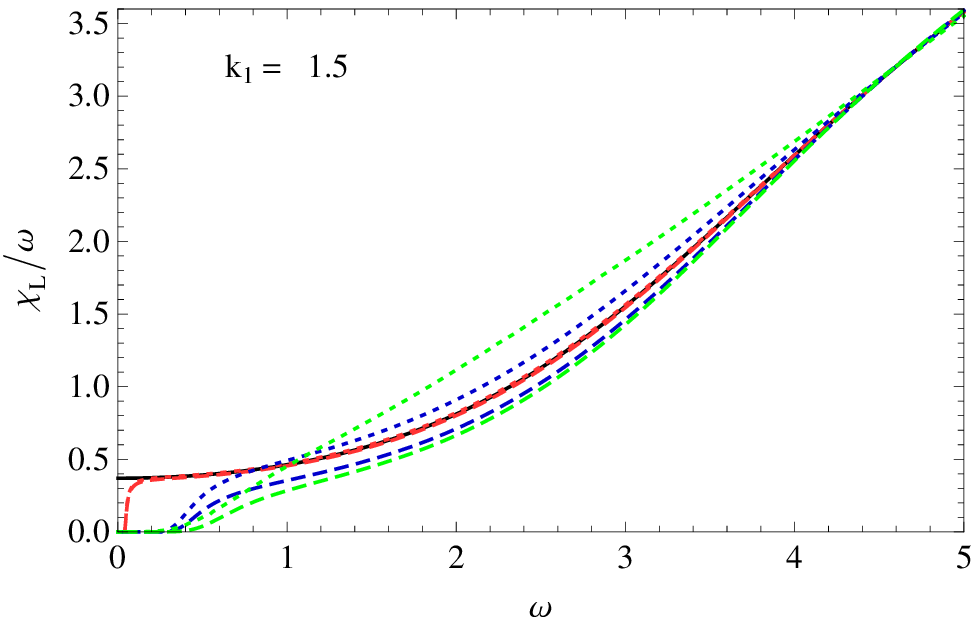}\quad%
\includegraphics[width=0.48\textwidth]{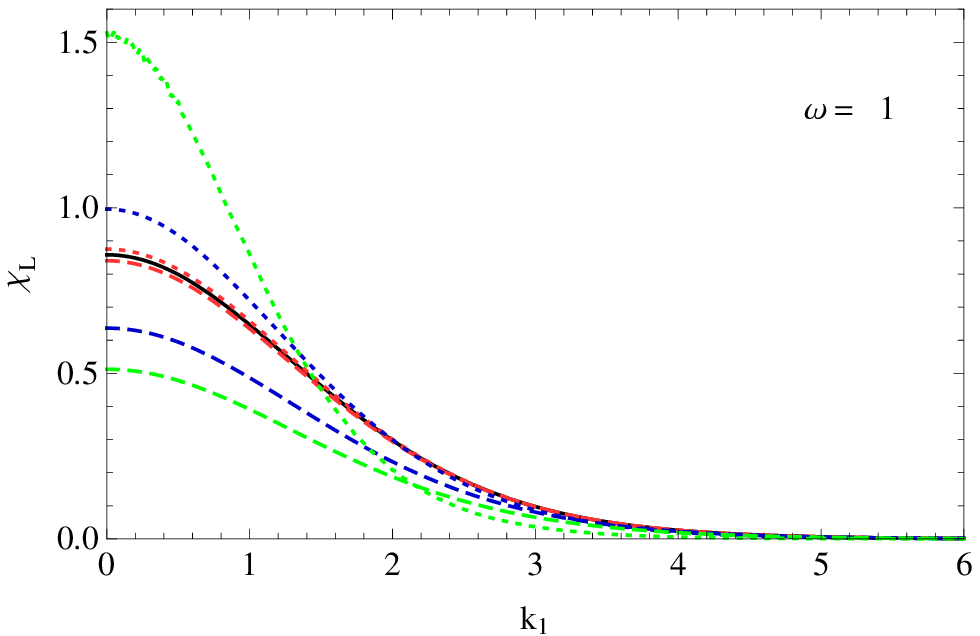}
\caption{Contribution to spectral function parallel to the anisotropy direction for $k_1$ fixed (left) and $\omega$ fixed (right). Color coding as in Fig. \protect\ref{fig:Light_TL}}
\label{fig:Momenta_LT}
 \end{center}
\end{figure}

For $\mathbf{k}=k_1\mathbf{e}_1$ there is another mode which is transverse with respect to the wave vector. The results for this mode, which is pointing along the anisotropy direction are shown in Fig. \ref{fig:Light_LT} and \ref{fig:Momenta_LT}. Compared to the previous modes that were transverse with respect to the wave vector, the dependence on $B$ changed. Here the spectral density is larger for prolate anisotropy and smaller in the oblate case. 

\begin{figure}
 \begin{center}
  \includegraphics[width=0.48\textwidth]{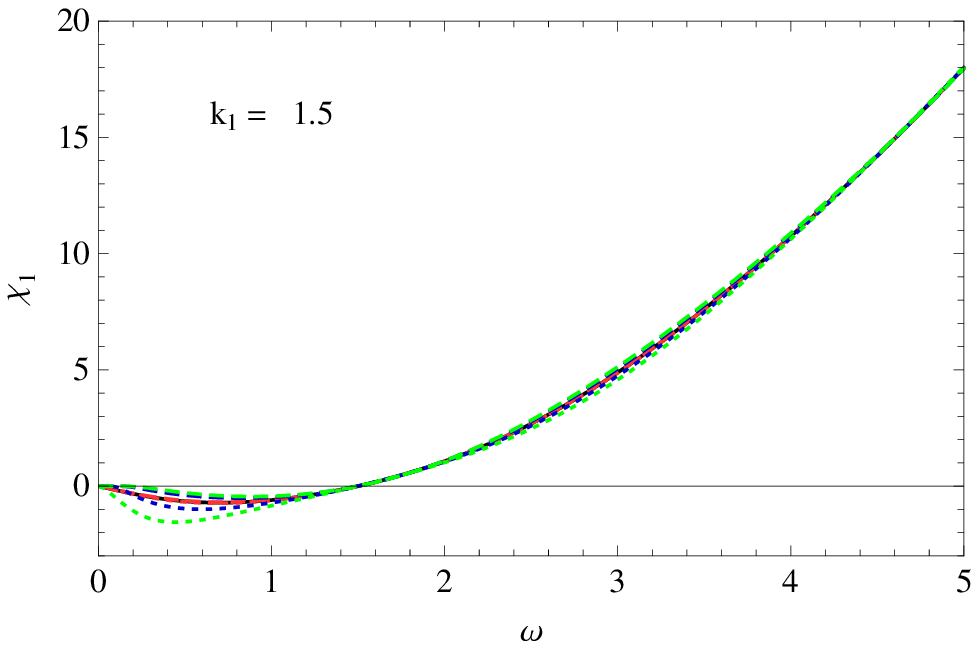}\quad%
\includegraphics[width=0.48\textwidth]{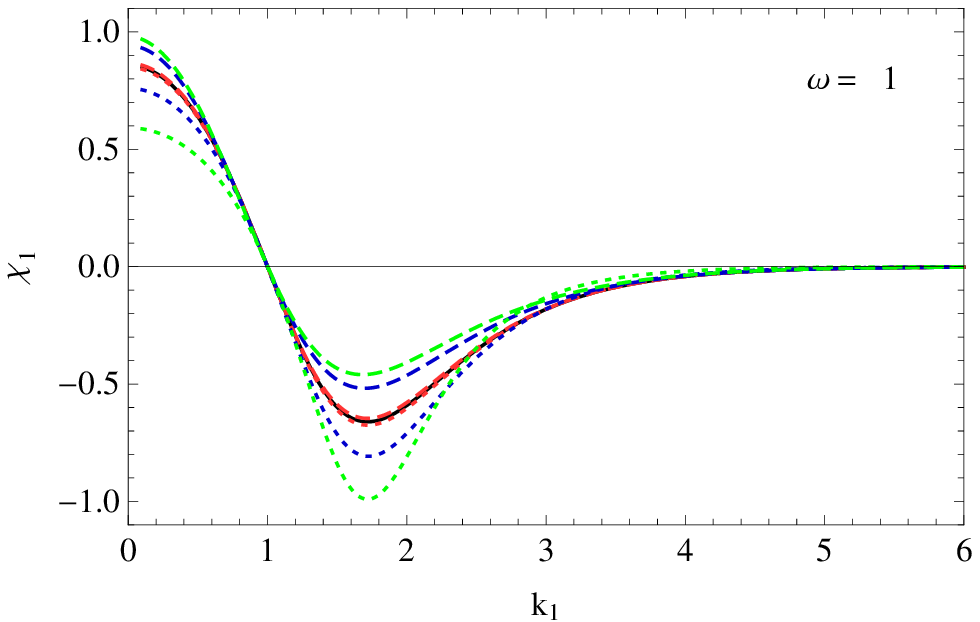}
\caption{Spectral function longitudinal to the wave vector for $k_1$ fixed (left) and $\omega$ fixed (right). Color coding as in Fig. \protect\ref{fig:Light_TL}}
\label{fig:Momenta_1T}
 \end{center}
\end{figure}

Finally we can consider the mode longitudinal to the wave vector. The behavior is shown in Fig. \ref{fig:Momenta_1T}. The spectral density for light-like momenta vanishes, as it should.

\subsection{Conductivities}

The fact that for any nonzero anisotropy parameter $B$ all spectral functions tend to zero stronger than linearly in the limit $\omega\rightarrow 0$ means that both the diffusion constant and the DC conductivity vanish according to Kubo's formulae. This absence of hydrodynamic behavior is clearly related to the absence of a horizon, which for $B\not=0$ gets replaced by a naked singularity.

\begin{figure}
 \begin{center}
\includegraphics[width=0.48\textwidth]{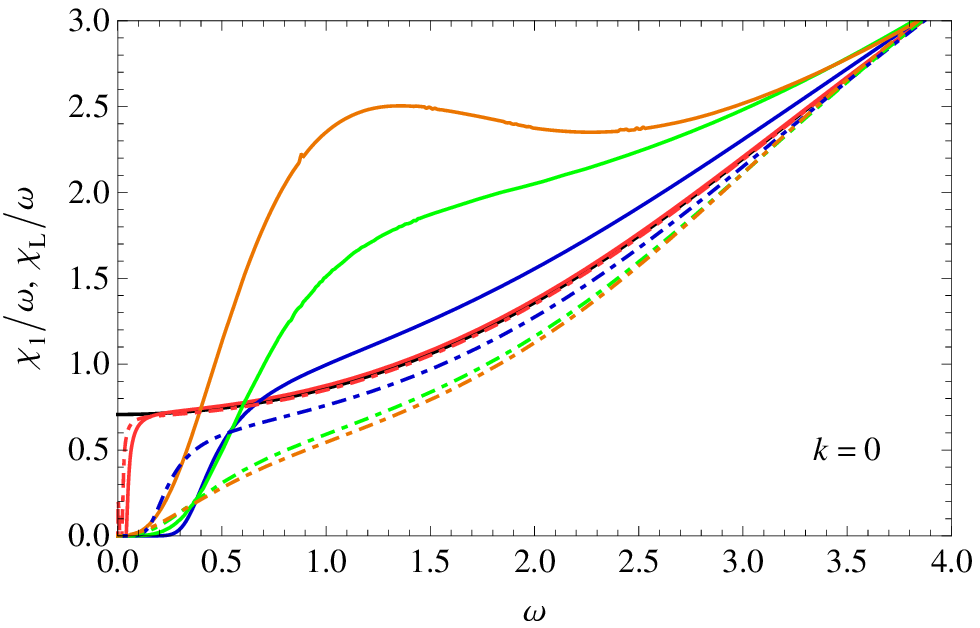}\quad%
\includegraphics[width=0.48\textwidth]{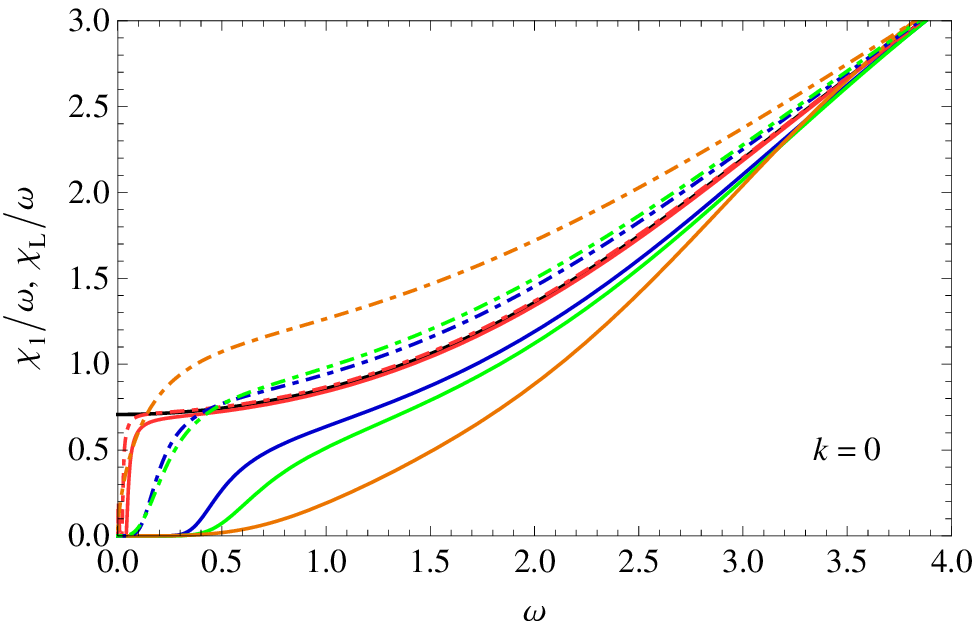}
\end{center}
\caption{Anisotropic AC conductivities: prolate vs. oblate anisotropies for various values of $B$. Full lines correspond to longitudinal conductivity,
dashed lines to transverse conductivity. Color coding of $B$ as in Fig. \protect\ref{fig:Light_TL}, with the addition of orange lines for $B=\mp3$, values for which negative pressures arise.}
\label{fig:ACcond}
\end{figure}

In Fig.~\ref{fig:ACcond} we display the results for the AC conductivities, juxtaposed for the cases of prolate ($B<0$) and oblate ($B>0$) anisotropies. In each case one can define longitudinal and transverse conductivities with respect to the direction of anisotropy. For prolate anisotropies transverse conductivities are found to be reduced compared to the isotropic case, whereas for oblate anisotropies this is true for longitudinal conductivities. However, in the limit of vanishing frequency, all conductivities go to zero. The frequency range in which this happens increases as the amount of anisotropy is increased, up to the point where one of the pressure components goes to zero. Curiously, when increasing the anisotropy parameter such that also negative pressures are produced, this trend is eventually reversed.

\subsection{Anisotropy of traced spectral function for light-like and time-like momenta}

\begin{figure}
 \begin{center}
\includegraphics[width=0.48\textwidth]{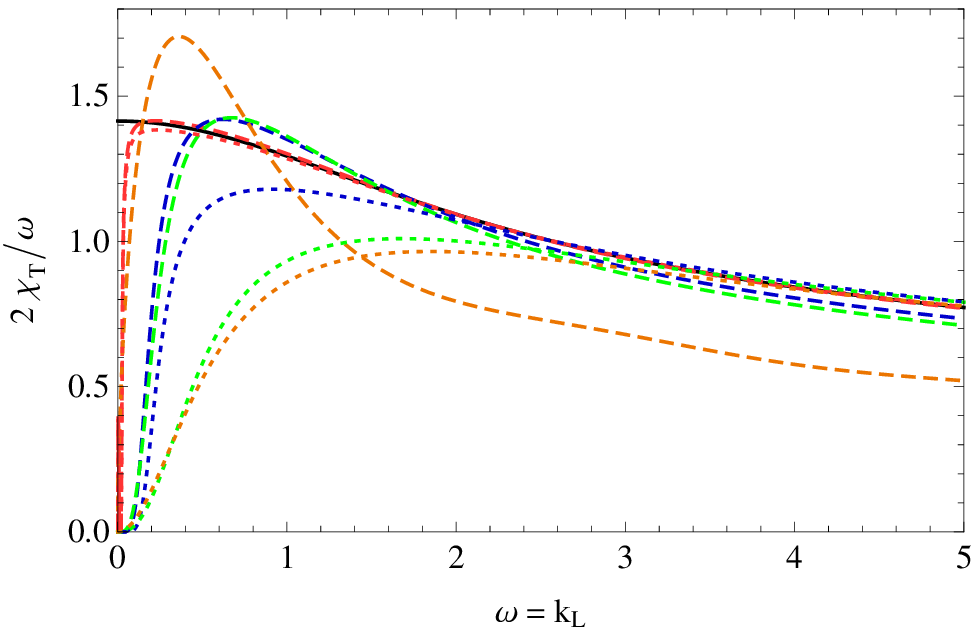}\quad%
\includegraphics[width=0.48\textwidth]{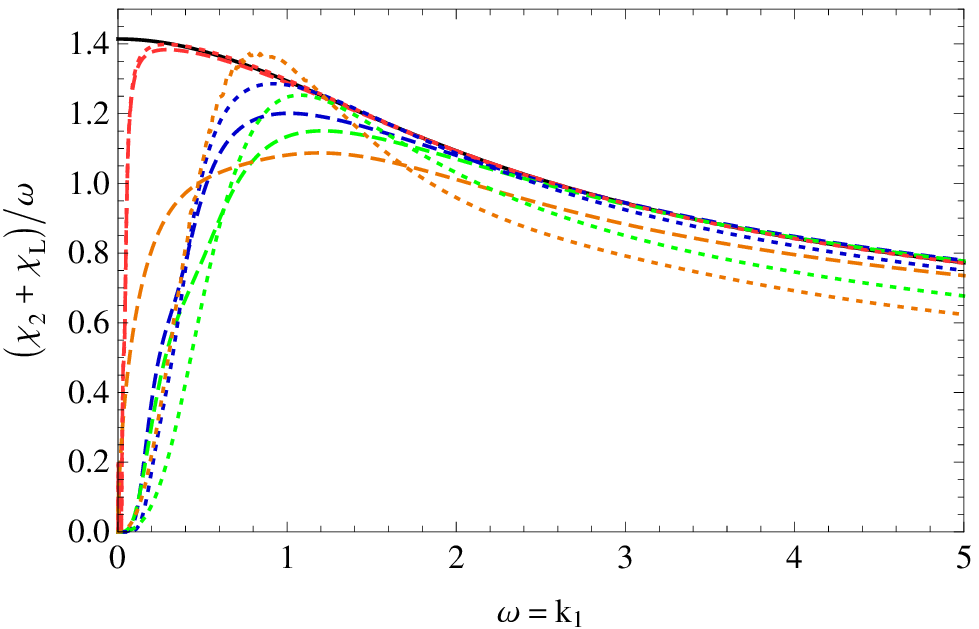}
\end{center}
\caption{$\chi^\mu{}_\mu/\omega$ for light-like momenta
with wave vector parallel to the anisotropy direction (left panel)
and transverse to it (right panel). Color coding of $B$ as in Fig. \protect\ref{fig:ACcond}}
\label{fig:comp-lightlike}
\end{figure}

\begin{figure}
 \begin{center}
\includegraphics[width=0.8\textwidth]{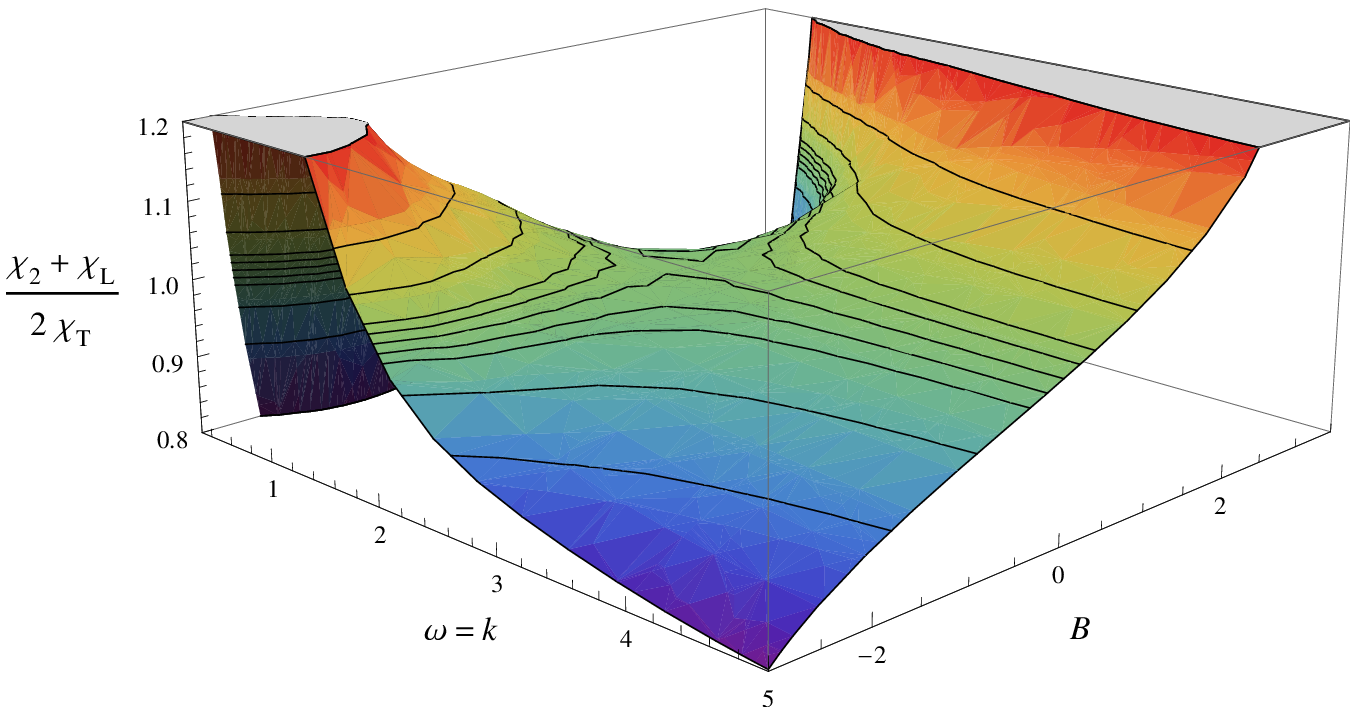}
\end{center}
\caption{Ratio of $\chi^\mu{}_\mu$ for light-like momenta with transverse wave vector over this quantity with parallel wave vector (w.r.t.\ the direction of anisotropy) as a function of both frequency and anisotropy parameter $B$.}
\label{fig:ratio-lightlike}
\end{figure}

The production of real photons is proportional to the trace of the spectral functions for light-like
momenta.
When the wave vector is parallel to the anisotropy direction, we have
\be
\chi^\mu{}_\mu(\mathbf k=k_L \mathbf e_L,K^2=0)=2\chi_T(k_L),
\ee
whereas for transverse wave vector, 
\be
\chi^\mu{}_\mu(\mathbf k=k_1 \mathbf e_1,K^2=0)=\chi_2(k_1)+\chi_L(k_1).
\ee
In Fig.~\ref{fig:comp-lightlike} the results for the wave vector pointing parallel and perpendicular to the anisotropy direction are juxtaposed, and in Fig.~\ref{fig:ratio-lightlike} the ratio between the latter and the former is shown as a function of frequency and anisotropy parameter $B$. 

For frequencies $\omega\lesssim 1$ (in units where $A=1$) this ratio shows
a rather dramatic dependence on the anisotropy parameter. However, as we have discussed above, we consider this regime to be unphysical since the assumption of stationarity of the anisotropic plasma ceases to make sense. For larger frequencies we indeed find a smoother dependence on the anisotropy parameter. For positive $B$ (oblate anisotropy) the spectral function with wave vector pointing in the direction of anisotropy is reduced, for negative $B$ the situation is reversed. As we shall discuss below, this is in line with the behavior of particle distributions for corresponding momentum anisotropies, but the latter typically have exponential suppression at high momentum.

\begin{figure}
 \begin{center}
  \includegraphics[width=0.48\textwidth]{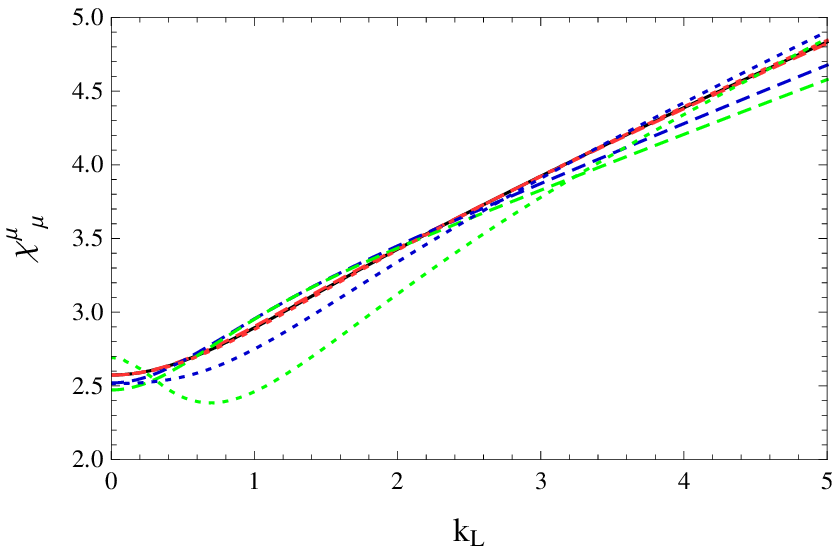}\quad%
\includegraphics[width=0.48\textwidth]{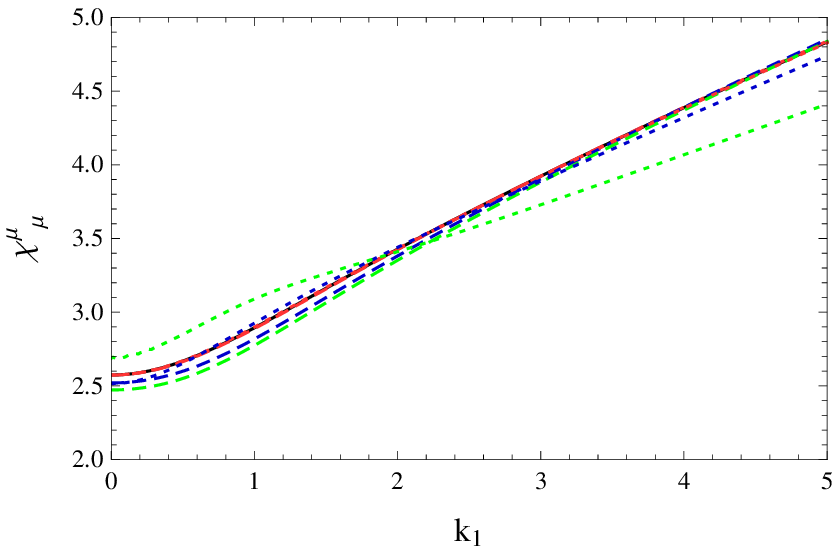}
\caption{$\chi^\mu{}_\mu$ for time-like momenta with invariant mass $K^2=1$ as a function of $k_L$ and $k_1$, respectively. Color coding as in Fig. \protect\ref{fig:Light_TL}}
\label{fig:dilepton}
 \end{center}
\end{figure}

In Fig.~\ref{fig:dilepton} we display the behavior of $\chi^\mu{}_\mu$ for time-like momenta, which is relevant for dilepton production, as a function of $k_L$ and $k_1$ for the two cases of longitudinal and transverse momentum at a fixed value of $K^2=1$.

\subsection{Photon and dilepton emission}

In order to obtain the photon and dilepton production rates, we must insert the results for the spectral function into (\ref{eq:photon production}) and (\ref{eq:dilepton production}), respectively. For the final result we would also need to know the distribution function $\phi$, which in the out-of-equilibrium situation is no longer fixed to the Bose-Einstein function. Absent the fluctuation-dissipation theorem, we should therefore calculate the 
Wightman function within the gauge/gravity duality framework directly. Attempts to incorporate the full formalism of nonequilibrium physics in the AdS/CFT correspondence were undertaken in \cite{Skenderis:2008dh, Skenderis:2008dg}, but we have not been able to apply these concepts to our case.

We shall instead make an estimate of photon and dilepton emission rates
by assuming that the distribution function $\phi$ is given by
the form (\ref{fiso}) with a parameter $\xi$ that at weak coupling
gives the same energy momentum tensor as in the boundary field theory of our gravity duality (see eq.~(\ref{Bvsxi})). 
Since the distribution function (\ref{fiso}) depends exponentially
on the angle,
the overall normalization constant 
$\mathcal{N}(\xi)$ does not play an important role. Because the energy
density at strong coupling, eq.\ (\ref{epsPLPT}), depends rather weakly on
the anisotropy parameter $B$, we have fixed $\mathcal{N}(\xi)$ such
that the energy density at weak coupling remains constant as $\xi$ and $B$
are varied. 

\begin{figure}
 \begin{center}
  \includegraphics[width=0.48\textwidth]{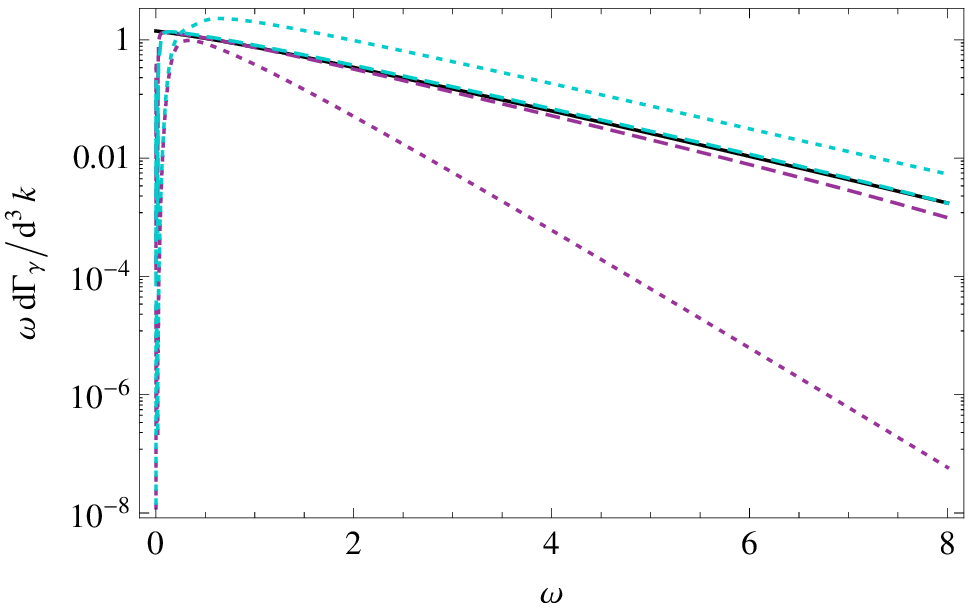}\quad%
\includegraphics[width=0.48\textwidth]{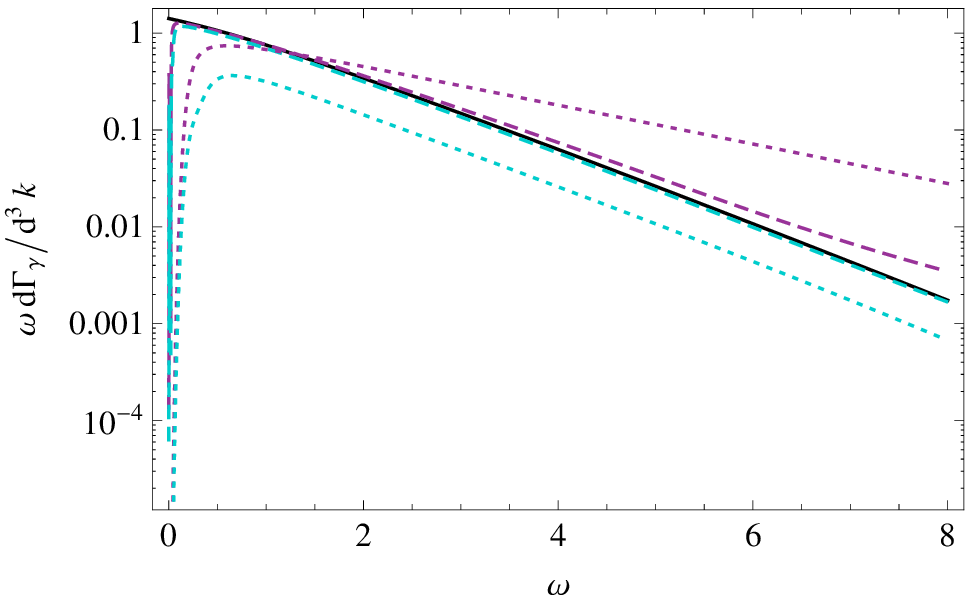}
\caption{Photon production rate in forward (purple) and transverse (cyan) direction for $B=\pm0.1$ (dashed) and $B=\pm1$ (dotted), where positive signs (oblate cases) correspond to the plot on the left and negative signs (prolate cases) are shown in the right panel. As a reference the isotropic rate is given in black.}
\label{fig:photo_1}
 \end{center}
\end{figure}

In Fig.~\ref{fig:photo_1} we display the resulting photon production rates
in longitudinal and transverse directions
for small and medium anisotropies. The left panel shows the situation
for oblate anisotropies ($B=0.1$ and 1), the right panel for
prolate anisotropies ($B=-0.1$ and $-1$).
This agrees qualitatively with the corresponding result for
oblate anisotropies in the weak coupling (hard-loop resummed) calculation of Ref.~\cite{Schenke:2006yp}.\footnote{Our estimated result for real photon production is in fact more strongly suppressed in the forward direction than the result of Ref.~\cite{Schenke:2006yp}. It shares this behavior with the soft part of the full result of Ref.~\cite{Schenke:2006yp}, which in the forward direction is dominated by hard contributions. The final result of Ref.~\cite{Schenke:2006yp} actually depends on the choice of a separation parameter of hard and soft scales which was fixed by a minimization procedure in the isotropic case. Fixing it instead anew for each value of $\xi$ would further reduce the hard-loop result in the forward direction.} Similar results are obtained for production rates of dileptons. The angular dependence in both cases is dominated by the function $\phi$ when it is chosen in accordance to the situation at weak coupling. 

\section{Conclusion}

In this work we have generalized the AdS/CFT calculation of
electromagnetic observables in strongly coupled $\mathcal N=4$
super-Yang-Mills theory in thermal equilibrium of Ref.\
\cite{CaronHuot:2006te} to the case of a stationary
anisotropic plasma using the singular geometry of
Janik and Witaszczyk \cite{Janik:2008tc}. In contrast to
the latter work, we have refrained from an expansion
in the anisotropy parameter since any nonzero anisotropy
changes qualitatively the character of the fluctuation equations.
In the limit of small frequencies, we have found that
for any amount of anisotropy the hydrodynamic behavior
of spectral functions is lost, whereas at higher frequencies
deviations from the isotropic case are smooth.
This reflects the fact that a stationary anisotropic geometry
can only be viewed as a reasonable approximation to
the early stage of plasma formation and thermalization 
when the involved time scales are sufficiently short. 

We have worked out the independent spectral functions of the
electromagnetic current-current correlation tensor
(defined by a weakly gauged U(1) subgroup of the R symmetry)
and their dependence on anisotropy parameter and wave vector.
Absent a recipe for directly calculating Wightman functions,
we have estimated photon production rates by adopting
deformed thermal distributions that have previously been employed
in weak coupling calculations. Doing so, our results turned
out to be qualitatively similar to the weak coupling
results of \cite{Schenke:2006yp} in that an oblate
anisotropy leads to a strong suppression in the forward direction.
It would of course be very desirable to have a direct
calculation of Wightman functions in our nonequilibrium
geometry which so far has been achieved only in few
cases \cite{Skenderis:2008dh,Skenderis:2008dg,CaronHuot:2011dr}.

Recently a different gravity dual for anisotropic plasmas has been
proposed in Refs.\ \cite{Mateos:2011ix,Mateos:2011tv} which is based
on geometries dual to Lifshitz-like fixed points constructed in
Ref.\ \cite{Azeyanagi:2009pr}. This approach involves additional
bulk fields and thereby achieves completely regular geometries.
It would be very interesting to compare spectral functions of current-current
correlators obtained in this approach with our results.
We would expect that the two approaches should agree in
the regime of high frequencies but deviate in the infrared.
However, in the latter regime neither of the two approaches will
provide a consistent model of a strongly coupled anisotropic plasma,
since the time dependence of the anisotropy can no longer be
ignored.

\subsection*{Acknowledgments}

We thank Rolf Baier, Karl Landsteiner, Bj\"orn Schenke, and Mike Strickland for
helpful discussions. This work was supported by the Austrian Science
Foundation FWF, project no. P22114, and \"OAD, project no. ES 12/2009.

\bibliographystyle{JHEP}
\bibliography{aniso}

\end{document}